\documentstyle[11pt,paspconf,epsf]{article}

\begin{document}
\def\gtsima{$\, \buildrel > \over \sim \,$}
\def\ltsima{$\, \buildrel < \over \sim \,$}
\def\simgt{\lower.5ex\hbox{\gtsima}}
\def\simlt{\lower.5ex\hbox{\ltsima}}
\def\smgt{$\simgt\,$}
\def\smlt{$\simlt\,$}
\def\smeq{$\simeq\,$}
\def\degs{\ifmmode^\circ\,\else$^\circ\,$\fi}
\def\kps{\ifmmode{\rm km}\,{\rm s}^{-1}\else km$\,$s$^{-1}$\fi}
\def\kms{\ifmmode{\rm km}\,{\rm s}^{-1}\else km$\,$s$^{-1}$\fi}
\def\ksmpc{\ifmmode{\rm km}\,{\rm s}^{-1}\,{\rm Mpc}^{-1}\else km$\,$s$^{-1}\,$Mpc$^{-1}$\fi}
\def\halpha{H$\alpha$}
\def\h1{\ifmmode h^{-1}\else$h^{-1}$\fi}
\def\onesigma{$1\,\sigma$}
\def\threesigma{$3\,\sigma$}
\def\nhat{\ifmmode {\hat{\bf n}}\else${\hat {\bf n}}$\fi}
\font\tensm=cmcsc10
\def\hii{H\kern 2.0pt{\tensm ii}}
\def\hi{\ifmmode{\rm H\kern 2.0pt{\tensm I}}\else H\kern 2.0pt{\tensm I}\fi}
\def\vtf{V_{{\rm TF}}}

\title{The LCO/Palomar $10,000$ km s$^{-1}$ Cluster Survey}

\author{Jeffrey A.\ Willick}
\affil{Physics Department, Stanford University, 
    Stanford, CA 94305-4060}

\begin{abstract}
I describe a Tully-Fisher (TF) survey of galaxies in 15 Abell clusters
distributed around the sky in the redshift range $\sim 9000$--12,000
km s$^{-1}.$ The observational
program was carried out during the period 1992--1995 at the
Las Campanas (LCO) and Palomar Observatories, and is
known as the LP10K survey. The data set consists of $R$-band CCD photometry  
and long-slit \halpha\
spectroscopy. The rotation curves (RCs) are characterized by
two parameters, a turnover radius $R_t$ and an asymptotic
velocity $V_a,$ while the surface brightness profiles
are characterized in terms of an effective exponential
surface brightness $I_e$ and scale length $R_e.$ The TF scatter
is minimized when the rotation velocity is measured at
$R=(2.0\pm 0.2)R_e$; significantly larger scatter results when
the rotation velocity is evaluated at $\simgt 3$ or
$\simlt 1.5$ scale lengths.
In contrast to most previous studies, a modest
but statistically significant surface-brightness
dependence of the TF relation is found, $\vtf\propto L^{0.28} I_e^{0.13},$
indicating a stronger parallel between
the TF relation and the corresponding Fundamental Plane relations
of elliptical galaxies than previously recognized.
 
To search for bulk flows on a 100 Mpc scale, a maximum-likelihood
analysis is applied to the LP10K data. The result is
$v_B=720 \pm 300\ \kms$ (\onesigma\ error)
in the direction $l=266\degs,$
$b=19\degs,$ with an overall \onesigma\ directional error of 38\degs.
This finding is in general agreement with the SMAC survey
of elliptical galaxies described in these proceedings.
However, it disagrees with recent findings on smaller scales
also described in these proceedings, notably those of Tonry
et al.\ using SBF data, Riess et al.\ from supernova 
data, the SHELLFLOW TF survey of Courteau et al, and
the SFI TF survey of Giovanelli and coworkers. 
The latter surveys convincingly demonstrate that the Hubble
flow has converged to the CMB frame by $cz \simeq 6000\ \kms.$
This suggests that the flows indicated by the SMAC and LP10K
data sets on $\sim 10,000\ \kms$ scales, which are of low
statistical significance in any case, should not be
taken literally. They result from a combination of
noisy data and, perhaps, local cluster peculiar velocities,
but do not represent bulk flow on a grand scale. 

\end{abstract}


\section{Introduction}

To understand the origins of the LP10K project, consider for
a moment where the field of large-scale flows stood in late 1990.
The ``Seven Samurai'' (7S) group (Lynden-Bell et al. 1988) had published
their then-stunning result that elliptical galaxies in the
nearby universe were streaming coherently in the direction
of the Great Attractor (GA) at $cz \approx 4000\ \kms,$
$l\approx 300\degs,$ $b \approx 10\degs.$ 
The effective depth of the 7S survey,
$\sim 3000\ \kms,$ was too small to know for sure whether the
motion was due simply to the gravitational pull of
the GA, or part of a much larger-scale bulk flow.
Hints that the latter might be the case came
from the large TF study of Mathewson et al.\ (1992),
who found galaxies in the GA itself to be outwardly streaming
like the 7S ellipticals more nearby, and 
from my own TF sample of Perseus-Pisces spirals
at $cz \approx 5000\ \kms,$
which appeared to be infalling on the opposite
side of the sky from the GA (Willick 1990).
It certainly seemed plausible in 1990 that bulk flows
might be coherent over much larger volumes
than the $\sim 5000\ \kms$ radius sphere probed
up to that time.

It was in this scientific atomosphere that I planned the LP10K study.
It's worth noting that I was unaware of the survey by Lauer
and Postman of brightest cluster galaxies (BCGs) that was then
nearing completion. Their work was, of course, soon to
burst on the scene with their 1992 announcement of a
high-amplitude ($\sim 700\ \kms$) bulk flow in which all
Abell clusters out to the huge distance of 15,000 \kms\ appeared
to partake (Lauer \& Postman 1994; LP). Subsequent to
the announcement of hte LP result, I often portrayed my
survey as ``testing Lauer-Postman'' for simplicity, but
in fact it was part of a larger picture with deeper roots.  

The LP10K survey is described in two journal articles (Willick 1999ab,
hereafter Papers I and II),
to which the interested reader should refer for technical
details. In this conference proceeding paper, I 
list some of the salient features of the survey, describe its
main results, and discuss how it fits into the picture
of cosmic flows that is taking shape as a result of
the Cosmic Flows 1999 conference.

\section{The LP10K Survey: Observations and Data Reduction}

The outlines of the survey took shape in
early 1991, when I was completing my PhD
thesis. 
I recognized then that to test for bulk flow one
needed {\em full-sky\/} data. This is not easy to come
by for a postdoc, but I had the good fortune to
become a Carnegie Postdoctoral Fellow.
Carnegie's unique resources
were its ample access to Palomar Observatory telescopes\footnote{Carnegie's
Palomar privileges ended in 1995, just as
my survey was completed.}
for Northern Hemisphere objects, and to telescopes at
the LCO for objects in the Southern sky. 

\begin{figure}[ht!]
\textwidth=4.25in
\plotone{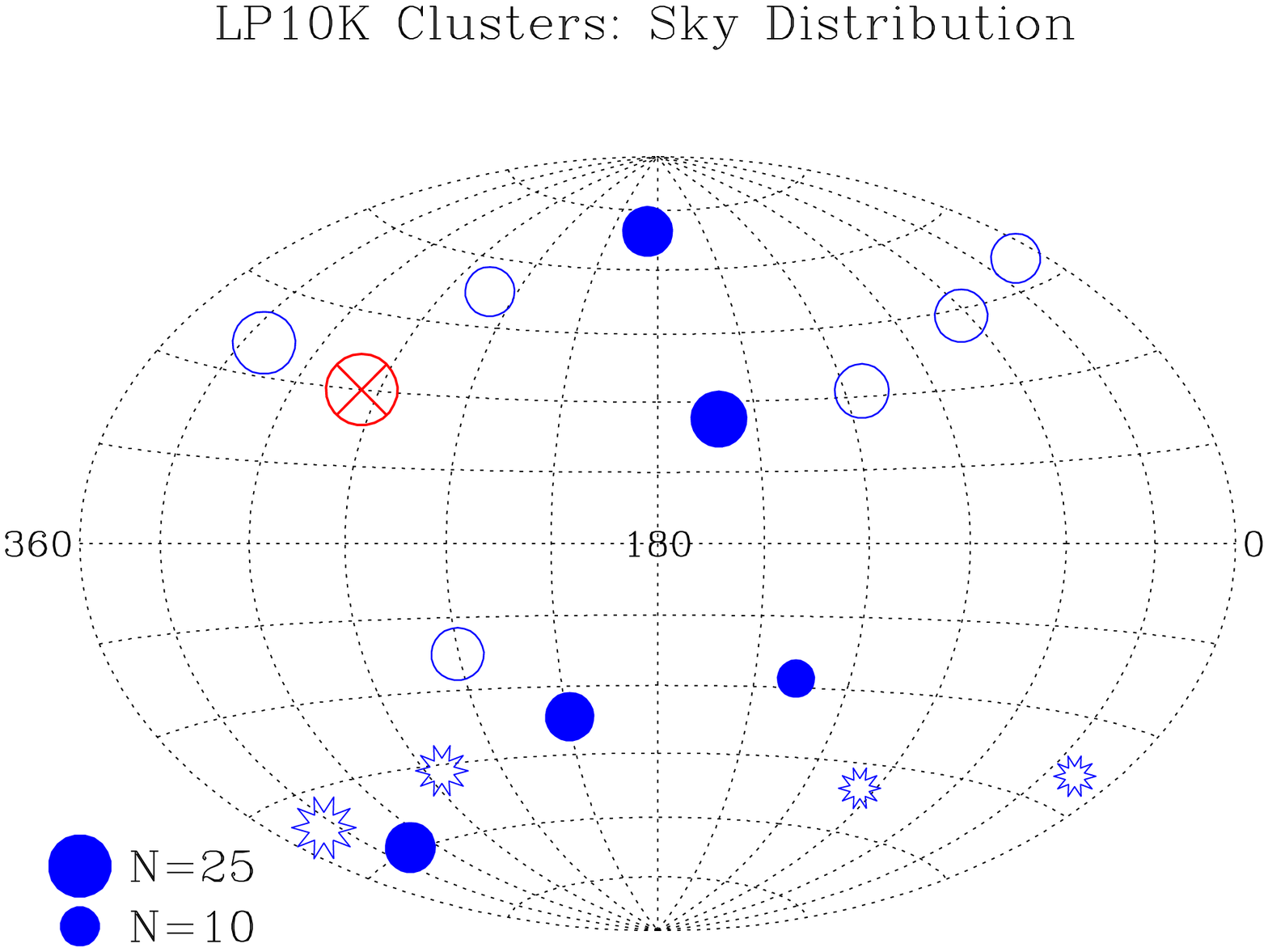}
\caption{Sky positions of the fifteen LP10K clusters.
Filled circles
represent clusters for which the mean redshift of
the LP10K TF sample is $\leq 12,\!000\ \kms.$
Open circles represent
clusters with mean TF sample redshift $12,\!000\ \kms \leq
cz \leq 15,\!000\ \kms,$ and starred symbols those
with mean $cz > 15,\!000\ \kms.$ The number
of galaxies in the cluster TF sample is symbolized by
the point size, as indicated by the key at the lower left.
The ``{\large $\times$}'' enclosed by a circle shows the direction of
the LG motion with respect to the CMB.
}
\label{fig:LP10Kdist}
\end{figure}

My intial plan, which I more or less followed, was to observe
galaxies in 15 Abell clusters distributed around the
sky in a relatively narrow redshift range, $9000 \le
cz \le 12,000\ \kms.$ I aimed to
get TF data for at least 15 galaxies in each cluster. 
The redshift slice was selected
so as to focus on a depth where there was little
or no extant data pertaining to cosmic flows. I searched
the ACO catalog (Abell, Corbin, \& Olowin 1989) for
clusters in this redshift range, and came up with about 35.
I then winnowed this list to 15 by randomly selecting an
``isotropic'' sample---that is, one that left no large
portion of sky unsampled, and which did not include
groups of clusters all in the same place on the sky.
Figure~\ref{fig:LP10Kdist} shows the positions of the
LP10K clusters on the sky, coding the size of the
TF samples and their median redshifts by the point type.

\textwidth=5.25in

In each
cluster I obtained $R$-band CCD imaging
of 1--2 square degrees of sky centered on
the cluster core. 
These moderately deep ($R \simlt 21.5$ mag) CCD frames
were then analyzed using FOCAS, and all galaxies
brighter than about $m_R=17.5$ were visually inspected
for suitability as TF galaxies.

The skeptical reader might pause here and ask what ``suitable''
means in this context.
Initially, it meant something rigorous
such as ``brighter than a certain apparent magnitude and
more inclined than a certain minimum inclination.''
I quickly learned that I could not afford to be so
picky, however. If I wanted galaxies whose rotation would
be detected, I needed
to pick galaxies that were emitting a lot of \halpha\ radiation.
The only way to ensure this was to choose objects primarily
on morphological appearance: they had to {\em look\/} they
had a lot of star formation going on. Basically, this
means well-defined spiral structure, generally late-type
morphology, and some visual evidence of \hii\ regions.
Moreover, I couldn't apply a single magnitude limit for
all clusters; rather, I had to go much deeper in clusters
with fewer bright, suitable spirals, in order to
obtain a sufficient number of TF galaxies.
This, I confess, is not a very rigorous selection
procedure, but it was the one that had to be
followed to make the program feasible.

Once the spirals were selected as described above, they
were observed spectroscopically at the LCO 2.5 m
or the Palomar 5 m telescopes. Long-slit spectra
were acquired at about 2 \AA\ resolution, in the
portion of the spectrum containing \halpha\ out to
$z=0.1$ Despite my efforts at preselecting objects
that would exhibit \halpha\ emission, I suffered
though many, many nondetections. This included objects
with nuclear \halpha\ emission, fine for redshift
purposes, but not extended enough to get a rotation
velocity. This should serve as a cautionary warning
to observers hoping to do deep, optical TF studies:
hone your selection criteria in advance so you don't
waste too much large telescope time. Detectable,
extended \halpha\ emission
is by no means guaranteed from faint spirals.

Because I selected TF galaxies from new CCD imaging and
not from a catalog, I didn't know in advance what their
redshifts would be, but I assumed they would usually be
close to the nominal values of their parent clusters.
This is where I got my second rude surprise (after nondetections):
a large fraction ($\sim 30$--$40\%$) of the TF sample
lay well in the background of the cluster. In the end,
fully two-thirds of the galaxies with extended \halpha\ emission
were found to have $cz > 15,000\ \kms,$ and a not insignificant
number had $cz \approx 30,000\ \kms.$ I intend at a later
date to explore the implications of these galaxies, which
are perfectly good Tully-Fisher objects, for monopole distortions to the
Hubble expansion, which is of considerable scientific interest. For
the immediate goal of bulk flow on a 100 Mpc scale,
of course, these objects were not especially useful.
In Figure~\ref{fig:LP10Kdist}, the point type codes the
median redshift of the TF sample. Note that the clusters
with the greatest preponderance of high-redshift objects
lie preferentially near the South Galactic Cap;
this result merits further attention, as it could be
indicative of very large-scale structure.

In the orginal survey plan,
I hoped to obtain Fundamental Plane (FP)
data for 15 ellipticals in each cluster to supplement, and provide
an independent check on, the TF distances. 
Indeed, an auxiliary scientific goal of the survey
was to use the spiral/elliptical cross-check as
a test for environmental effects---significant
E/S distance discrepancies that correlated with
cluster properties would provide evidence for these.
Unfortunately, the elliptical galaxy spectroscopic data is still
very much in the reduction process, in the case
of LCO ellipticals which I could observe using
the du Pont telescope multifiber spectrograph, and
are still to be acquired in the case of Northern
sky ellipticals. If the LP10K
elliptical results ever come out, it will probably be
via a healthy amount of merging with other data sets.
In hindsight, I can see that including ellipticals
in the survey plan was a proverbial case of ``biting
off more than I could chew.'' I say this not 
for the purpose of public self-flagellation but as a cautionary tale 
for ambitious postdocs contemplating a massive
observational program. In any case, readers should
keep in mind that
the results presented
in Papers I and II, and here, are based on 
Tully-Fisher data only.

\section{Formulating the Optical TF Relation}

\subsection{Defining the TF Rotation Velocity}
A key question that the LP10K survey data addressed
was, ``Given that we have rotation curve (RC)
data, $V(R),$ for each galaxy, what particular velocity
$\vtf$ is it that enters into the TF relation?''.
While similar
questions arose in the older, \hi-based
TF surveys, where one had to extract a width
from an unresolved 21 cm profile, the problem then
was largely algorithmic. 
With long-slit
spectroscopy, however, we have resolved data, 
and the issue of ``{\em what\/} rotation velocity enters
into the TF relation'' becomes a physically meaningful one.

For LP10K I took the following approach. First, the
RCs were fitted using a two-paremeter functional form:
\begin{displaymath}
V(R) = \frac{2 V_a}{\pi} \tan^{-1} \left(\frac{R}{R_t}\right)\,.
\label{eq:varctan}
\end{displaymath}
The arctangent form approximates the canonical S-shaped rotation
curve, but can also represent the quasi-linear RCs, still
rising at the outermost point, that are frequently encountered.
It is useful to refer to $V_a$ as the ``asymptotic'' velocity,
and to $R_t$ as the RC scale
length or ``turnover
radius.'' It is important to bear in mind, however, that
the velocity $V_a$ is not necessarily reached by the sampled
RC, and may in fact not be reached at all, so its name, while
evocative, is not rigorous. Note also that for a quasi-linear
RC, only the ratio $V_a/R_t$ is well-determined from the fit.
Still, bearing these caveats in mind, the arctan fit proved
quite adequate for all LP10K galaxies. 

The photometric data also yields a scale length for the
galaxy, viz., $R_e,$ in this case the effective exponential
scale length (see Paper I for more detail on how this 
was computed; it was {\em not\/} from an exponential fit).
With it, one can pose the question of ``what is $\vtf$''
in this way: at how many exponential scale lengths
should one evaluate the rotation curve $V(R)$ to
get the best (lowest-scatter) TF relation? Let us
parametrize this question by writing $\vtf=V(f_s R_e),$
where the right side is evaluated from the arctan fit,
and take $f_s$ as a free paramer.
I like to write this in the following way:
\begin{displaymath}
\vtf = \frac{2 V_a}{\pi} \tan^{-1} \left(\frac{f_s}{x_t}\right)\,,
\label{eq:vxt}
\end{displaymath}
where $x_t \equiv R_t/R_e$ is a dimensionless shaper parameter
which measures the ratio of the dynamical to luminous scale lengths
of the galaxy. Roughly speaking, $x_t \ll 1$ is a classical
S-shaped RC, with the flat part well sampled, while $x_t \simgt 1$
is a quasi-linear RC. 

\begin{figure}[ht!]
\textwidth=4.25in
\plotone{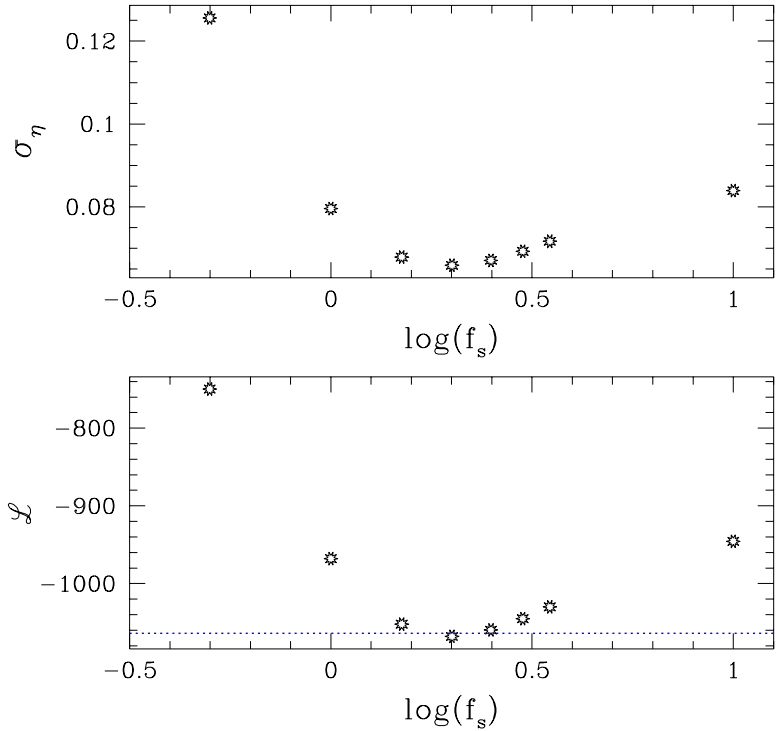}
\caption{The inverse TF scatter (top panel) and 
the fit likelihood statistic, ${\cal L}=-2\ln P({\rm data}|{\rm model}),$
plotted versus the value of $f_s.$ Both plots show a clear
minimum (i.e., a best fit) at $f_s=2.0\pm 0.1.$}
\label{fig:fsl}
\end{figure}
\textwidth=5.25in

We next define the velocity width parameter, in the
usual way, $\eta(f_s)=\log(2\vtf)-2.5,$ and write the
inverse TF relation $\eta(f_s)=-e(M-D).$ This is a three-parameter
TF relation, with slope $e$ and zero point $D$ supplemented
by $f_s.$ The values of all the parameters can be determined
by maximum likelihood (see Papers I and II), 
with absolute magnitude given by a simple Hubble Flow
model (in the CMB frame). It is illustrative to do
this for a range of fixed values of $f_s,$ maximizing
likelihood with respect to $e$ and $D$ at each value
and obtaining a correponsponding TF scatter and fit likelihood.
The results of this exercise are presented in Figure~\ref{fig:fsl}.
The plot shows that the ``correct'' value of $f_s$---i.e.,
the one that minizes TF scatter---is quite tightly
constrained at $f_s\approx 2.0.$ In particular, one sees
that taking $f_s \gg 1,$ corresponding to taking $\vtf=V_a,$
leads to a poor TF relation. This is an important point,
for it demonstrates that the canonical wisdom---that
the TF relation involves the velocity on the flat part
of the RC---is wrong. In Paper I a heuristic physical
explanation for this effect is offered, but undoubtedly
the correct explanation is much deeper, and I encourage
theorists to delve more deeply into this issue. 

\subsection{A Surface Brightness Dependence of the TF Relation}

Another galaxy property which turned out to enter
into the TF relation was the effective surface
brightness $I_e$ (calculated via the same method
that produced the scale length $R_e$---see Paper I
for details). Specifically, we can write the TF
relation
\begin{displaymath}
\eta(f_s) = -e (M-D) - \alpha \mu_e \,,
\label{eq:4parmTF}
\end{displaymath}
where $\mu_e$ is the magnitude equivalent of
$I_e.$ When this was done, a small but
statistically significant reduction
in the TF scatter was found; the coefficient
was found to be $\alpha \approx 0.05,$ which when
combined with $e\approx 0.12$ leads to a power-law
scaling relation of the form $\vtf \propto I_e^{0.13} L^{0.3}.$
This relation is reminiscent of the FP relations for
elliptical galaxies. The surface brightness dependence
of the TF relation has significant implications for
galaxy structure.

It is only fair to note here that my claim of an SB
dependence of the TF relation is not universally accepted.
Riccardo Giovanelli made the very good point at the conference
that by introducing a scale length into the definition of
velocity width, I may have induced a spurious SB dependence
of the TF relation. My Shellflow colleague St\'ephane Courteau
has argued that the particular method I employ for
determining $R_e$ and $I_e$ from the surface brightness
profiles could produce the dependence, whereas in a more standard
approach the dependence would not show up. As of this
writing (8/31/99) these possibilities have not been
fully investigated. I hope to study this issue more closely in the coming
months with the Shellflow data, and in the meantime I encourage
input from members of the community who have done such
experiments with their own data.

\subsection{The Intrinsic TF Scatter}

The likelihood fits used in Papers I and II made it possible
to constrain the individual contributions to the overall TF scatter: 
measurement errors and intrinsic scatter. 
Although the rotation 
velocity measurement errors were not well determined
a priori, they can be separated from the intrinsic
scatter because the TF error they induce scales as
$\delta\eta \propto \delta v/v.$
(Photometric errors are fairly well determined, and
are small in any case.)
Thus, if one assumes a fixed velocity measurement error, 
the overall $\eta$ error $\propto \vtf^{-1}.$ Using this
model, I found that (i) the characteristic rotation velocity
measurement error was $\sim 17\ \kms,$ a reasonable value
consistent with repeat observation comparisons;  
(ii) the observed
decrease of TF scatter with increasing luminosity
can be fully accounted for in this way---i.e., 
we are not {\em required\/} to assume the intrinsic
TF scatter decreases with increasing luminosity---and
(ii) the intrinsic TF scatter itself is $0.28 \pm 0.07$ mag.
This value is consistent with what was found by Willick et al.\ (1996)
via an independent approach, and is also consistent
with findings from the Shellflow survey (Courteau et al., this
volume). Reproducing this level of scatter is an important
challenge for galaxy formation theory.

\section{On Bulk Flows }

To determine the best-fitting bulk flow vector
$\vec v,$ I calculated absolute magnitude as
$M=m-5\log[(cz-\vec v \cdot \nhat)/H_0]-25,$ and then maximized
TF likelihood as before. Paper II gives many details about
these fits, but the upshot is that the derived 
flow vector has an amplitude of $\sim 700\pm 300\ \kms$
in the direction quoted in the Abstract. The relevant
question now is, what should we make of this? As
those who heard my talk at the conference, or heard
about it, already know, I no longer consider this result to be ``correct.''

Unfortunately, this seems to have led to a perception 
that I am ``disavowing my own data.''
This is untrue. The data is perfectly good, or at least
as good as it can be given the observational challenges
described above. Rather, my point of view is that
this result is noisy---the flow is detected at
the $\sim 2.3\sigma$ level---and thus would require
extensive corroboration to be accepted. 
Looking at the other data sets around---most of them
presented at the Cosmic Flows workshop---the prevailing
picture is one of convergence to the CMB frame on smaller
scales. Most convincing to me, because I worked on it,
are the Shellflow results, which clearly show convergence
at 6000 \kms. Also very persuasive in this regard are the new supernova
results presented by Adam Riess.

To believe the LP10K bulk flow result, then, I would have
to accept that the Hubble flow converges to the CMB frame
at 6000 \kms, and then starts drifting away from the
CMB frame beyond this distance. This would be blatantly
unphysical. By and large I harbor few prejudices about
how the universe should behave, but I basically do 
believe that gravitational instability drives structure
formation and flows, and that the universe 
approaches homogeneity on progressively larger scales.
If this is true, and Shellflow and the other
shallower surveys are correct, the LP10K flow result
cannot be. That is not a disavowal of my own data,
but quite basic and, I hope, sound scientific reasoning.

\acknowledgments
First and foremost, I once again thank my friend and
collaborator St\'ephane Courteau for organizing this
fantastic conference. I also thank Felicia Tam,
a Stanford undergraduate, for invaluable assistance
in reducing the LP10K data over the last two years.

\end{document}